# Size-dependent orbital symmetry of hole ground states in CdS nanocrystals

Boqian Yang,[1,2] James E. Schneeloch,[1] Zhenwen Pan,[3] Madalina Furis,[3] Marc Achermann[1,*]

[1]Physics Department, University of Massachusetts Amherst, Amherst, MA 01003, USA.
[2]Physics Department, University of Puerto Rico, Rio Piedras, PR, 00931, USA.
[3]Physics Department, University of Vermont, Burlington, VT 05405, USA.

(Dated: October 13, 2009)

Using time-resolved photoluminescence spectroscopy, we studied the electronic levels of semiconductor nanocrystals (NCs) with small spin-orbit coupling such as CdS. Low temperature radiative rates indicate that the lowest energy transition changes from orbital allowed to orbital forbidden with decreasing NC size. Our results are well explained by a size-dependent hierarchy of s- and p-orbital hole levels that is in agreement with theoretical predictions. Around the critical NC radius of ~2 nm, we observe an anti-crossing of s- and p-orbital hole levels and large changes in transition rates.

PACS numbers: 73.22.-f, 78.47.-p, 78.55.Et, 78.67.Bf

Size-tunability of optical properties and flexibility of surface chemistry make semiconductor nanocrystals (NCs)[1, 2] attractive fluorophores for a wide range of applications including solid-state lighting [3, 4], lasing [5, 6], and fluorescent labeling [7, 8]. High emission efficiencies with quantum yields larger than 50% were demonstrated [9-11], despite the optically passive character of the lowest energy transition in many NCs [12-20]. The high emission yields at room temperature are obtained because of thermally populated excited transitions that are characterized by high oscillator strengths. Understanding such optical properties of NCs requires a detailed knowledge of their electronic level structure. Extensive studies of CdSe NCs led to the concept of dark and bright excitons [12-15]. Significantly less experimental studies exist on the electronic level structure of NCs made of other semiconductors, including CdS [16-18]. Besides the application-relevant emission in the blue and UV spectral range [21-23], the importance of CdS NCs arises from the distinct differences in the electronic band structure between CdS and CdSe. Specifically, CdS is a representative of semiconductors (e.g. sulphides and phosphides) with very small spin-orbit coupling $\Delta_{SO}$ in contrast to CdSe which has a large spin-orbit coupling.

In this report, we discuss the electronic level structure CdS NCs (as an example of a small $\Delta_{SO}$ material system) using time-resolved photoluminescence (PL) measurements. We determined temperature- and size-dependent radiative decay rates of CdS NCs and analyzed them with a model that takes into account two lowest energy transitions between s-orbital electron and s- or p-orbital hole levels. We demonstrate for the first time experimental evidence that the orbital symmetry of the hole ground state in CdS NCs depends on the NC size, a phenomenon that has only been predicted theoretically [24, 25]. Around the critical NC size, our measurements indicate an anti-crossing of the s- and p-orbital hole levels that is accompanied by significant changes in the transition rates. Our work highlights the interesting regime of intermediate CdS sizes and bridges the gap between a few reports on small and large CdS NCs [16-18, 23]. Moreover, the results presented here are an important contribution to the controversial debate about the appropriate theoretical modeling of such NC materials [24, 26, 27].

The electron and hole level structure of II-IV NCs that results in their characteristic absorption and emission properties has been studied intensively in the last years. Common to many NCs is the relatively large Stokes shift between absorption and emission energies [13, 18]. A good understanding of this effect has been provided by calculations based on the effective-mass approximation [13]. For CdSe NCs, for example, NC shape anisotropy, intrinsic crystal field, and electron-hole exchange interaction lift the eight-fold degeneracy of the exciton ground state resulting in a five-level exciton "fine structure" [13]. According to their total angular momentum projection, the five levels can be devised into three dipole allowed (bright) and two dipole forbidden (dark) exciton levels that give rise to the observed *global* Stokes shift between absorption and emission spectra and the *resonant* Stokes shift between the ground state dark exciton and the first excited state bright exciton [12-15, 28].

Since the ground state transition in CdSe NCs is *spin* forbidden, radiative lifetimes measured at low temperatures are very long, in the μs range [12, 15]. Similarly, long low-temperature radiative lifetimes were found in most other II-IV NCs [18-20]. While the excitonic fine structure needs to be considered for explaining such emission properties of *CdSe* NCs, many optical properties of *CdS* NCs can be described without taking into account electron-hole exchange interactions. In small CdS NCs, calculations of electron and hole levels have shown that the lowest energy electron-hole transition is *orbital* forbidden because of s-orbital electron and p-orbital hole ground states. The next higher-energy transition involves s-orbital hole levels and is *orbital* allowed [24, 25, 29-31]. This electronic level structure has successfully been used to explain the resonant Stokes shift [18]. Interestingly, it has also been predicted that the hole level hierarchy is not universal in CdS NCs, but depends on the NC size such that the s-orbital hole levels become ground states for large CdS NCs [24, 25]. Hence, the lowest energy transition is expected to be allowed in large CdS NCs.

To assess the orbital symmetry of the lowest energy hole levels in CdS NCs, we studied four different sizes of CdS NCs that were produced by NN-labs following the procedure in Ref. 33. The CdS NCs are stabilized by oleic acid ligands, dissolved in toluene, and show size tunable room-temperature (RT) emission with quantum yields in the 5-10 % range and peaks at 415, 438, 452, and 472 nm [Fig. 1(a)]. Absorption measurements reveal 1S peaks at 399, 424, 440, and 459 nm [Fig. 1(b)]. The NC radii of 1.7, 2.1, 2.4, and 2.7 nm, respectively, were reported by NN-labs based on TEM imaging. Throughout this report the indicated NC size will be the radius. For time-resolved photoluminescence (PL) measurements, we excite the NCs with a low-power diode laser providing ~50 ps pulses at a wavelength of 407 nm. The luminescence is spectrally and temporally resolved by a monochromator equipped with a time correlated single photon counting system (TCSPC) that yields a time resolution of 70 ps. Excitation or our smallest NC sample leads to a size selection effect and, consequently, the probed NCs have a slightly larger radius of 1.8 nm (instead of 1.7 nm). Low temperature (LT) studies were performed with a continuous flow liquid-He cryostat that allows for temperature control in the range 4-300 K. While RT measurements were performed with solution samples, we used drop-cast films of just NCs or NCs in a PMMA matrix for temperature dependent measurements.

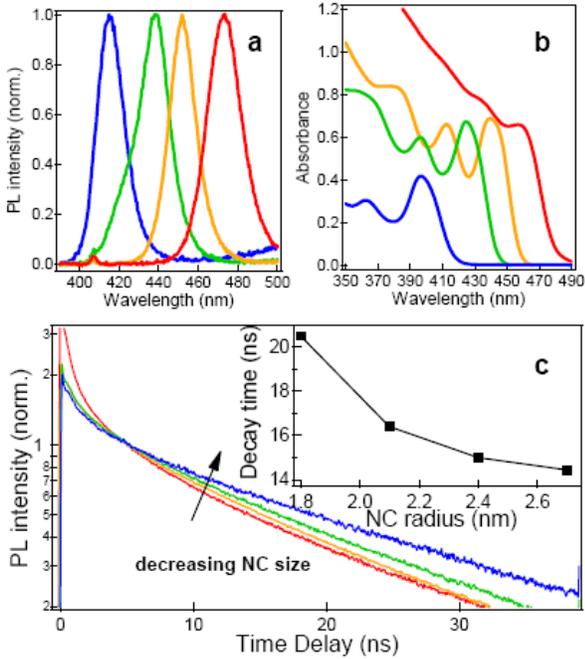

**Fig. 1:** Band edge emission (a) and absorption (b) spectra of the four CdS NC samples. (c) Room temperature PL dynamics; normalized at a time delay of 5 ns. Inset: Lifetimes as a function of CdS NC radii. Color code for entire figure: blue: 1.8 nm NCs; green: 2.1 nm NCs; orange: 2.4 nm NCs; red: 2.7 nm NCs.

In Fig. 1(c) we show RT PL dynamics taken at the peak emission wavelength of the four CdS NC solutions. After an initial fast decay the PL dynamics of all samples show slow, nearly single exponential decays. Such PL decays on multiple time scales are typical for measurements of NC ensembles that have quantum yields of less than unity and, therefore, contain subsets of NCs with non-radiative and radiative decay characteristics. The single exponential PL decay at long time delays indicates that a subset of excited NCs decays predominantly radiatively [15]. We determine the radiative lifetimes, $\tau_{rad}$, by fitting the experimental data at time delays > 10 ns with a single exponential function. The obtained $\tau_{rad}$ are in the range 14 - 21 ns and demonstrate that longer lifetimes correspond to smaller NCs [inset of Fig. 1(c)]. The extrapolation of our results to very small NC sizes is consistent with previous work that reported $\tau_{rad}$ = 40 ns for 1.1 nm CdS NCs [18].

The long $\tau_{rad}$ at room temperature have been explained by taking into account both the lowest-energy *forbidden* transition between s-orbital electron and p-orbital hole levels and the first excited *allowed* transition that is thermally populated and involves s-orbital hole levels [18]. As mentioned before, we do not take into account electron-hole exchange interactions in our discussion here; hence, the electronic level assignments refer to electron and hole levels with either s- or p-orbital character and allowed or forbidden transitions refer to orbital selection rules without accounting for spin (in contrast to the context of dark/bright excitons in CdSe). Since the energy difference between the two allowed and forbidden transitions (often referred to as resonant Stokes shift [18]) decreases with increasing NC size, a reduction of $\tau_{rad}$ is expected for larger NC sizes, consistent with our measurements.

For further insight into the details of the two lowest energy transitions in CdS NCs, we performed temperature dependent PL decay measurements. At low temperatures thermal population of excited states is unlikely and, therefore, the emission originates primarily from the ground state transition. According to the simple explanation introduced above, one would expect a deceleration of the PL dynamics, because of the orbital forbidden s-electron / p-hole ground state transition. In Fig. 2(a) we compare PL decays at T ~ 4K and RT of the 1.8 nm and the 2.7 nm CdS NCs. Interestingly, the PL dynamics indicate different temperature dependences for the two sizes. For the small, 1.8 nm NCs, the decay slows down at lower temperatures, as expected, and consistent with results obtained for 1.1 nm CdS NCs[18] and CdSe NCs [12, 15]. From single exponential fits at time delays > 10 ns we determine $\tau_{rad}$ = 33 ns and $\tau_{rad}$ = 21 ns at low and room temperature, respectively. The situation for the 2.7 nm CdS NCs is opposite; we measure an acceleration of the PL decay at low temperatures (LT and RT lifetimes of 7 ns and 14 ns, respectively). This temperature dependence of the PL dynamics in large NCs is unusual and unexpected and only few studies have reported similar results [23]. Such a behavior can be explained by a lowest energy *allowed* transition in large CdS NCs, in contrast to the *forbidden* transition in small CdS NCs.

In Fig. 2(b) we show $\tau_{rad}$ obtained from a large set of PL decay measurements as a function of temperature and CdS NC sizes. It is noteworthy that the RT radiative lifetimes in this data set obtained from NC *film* samples are comparable to results in Fig. 1(c) from NC *solution* samples; therefore, we conclude that film effects such as energy transfer between NCs are negligible in the analyzed data [32]. The temperature dependent lifetimes of each NC sample indicate that the transition between LT and RT values is not a gradual change, but occurs around a specific transition temperature that depends on the NC size and is larger than ~70 K for all NC sizes.

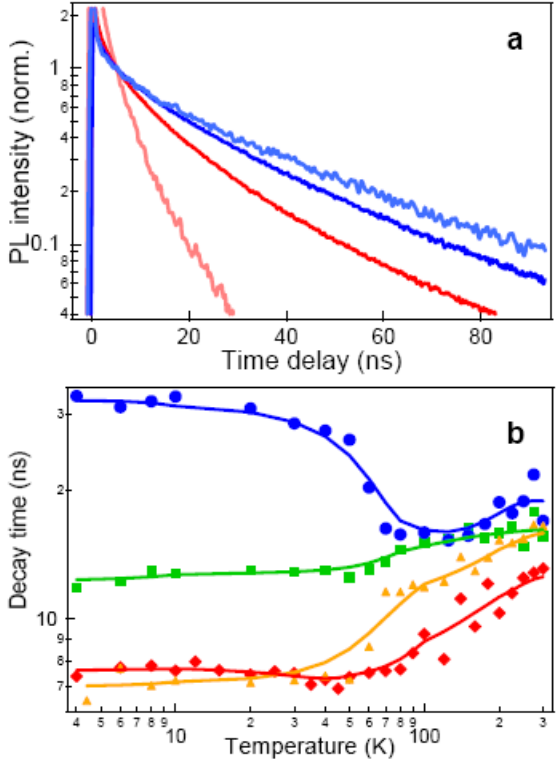

**Fig. 2:** (a) PL dynamics of the 1.8 nm and 2.7 nm NCs taken at RT and ~4 K (red: 2.7 nm NCs at RT; light red: 2.7 nm NCs at LT; blue: 1.8 nm NCs at RT; light blue: 1.8 nm NCs at LT). (b) Temperature dependent PL decay times. (blue circles: 1.8 nm NCs; green squares: 2.1 nm NCs; orange triangles: 2.4 nm NCs; red diamonds: 2.7 nm NCs). Solid lines are guide for the eyes.

Most importantly, we observe different temperature dependences of $\tau_{rad}$ for different NC sizes that point out that the hierarchy between forbidden and allowed transitions depends on the NC size. The 1.8 nm CdS NCs have temperature-dependent $\tau_{rad}$ that display the characteristics of a lowest-energy forbidden transition and first-excited allowed transition [15]. In contrast, the temperature dependent $\tau_{rad}$ of 2.4 nm or larger CdS NCs indicate a lowest-energy allowed transition that leads to relatively short $\tau_{rad}$ at LT and a first-excited forbidden transition that is thermally populated at RT and, therefore, causes a prolongation of $\tau_{rad}$. Such a size-dependent switching in the hierarchy of forbidden and allowed transitions has been predicted theoretically, supporting our assertion [24, 25]. These calculations indicated that the s (p) state is the ground hole level for CdS sizes larger (smaller) than 6.9 nm [25] or 4.5 nm [24]. Though both theoretical papers confirm the general explanation of our measurements, the difference in the critical NC size, at which the s- and p-orbital hole levels change their order, can be explained by the approximation of theoretical models that did not take into account coupling between valence and conduction bands [29] and the sensitivity of the calculations to material parameters.

For a quantitative discussion of $\tau_{rad}$ and the electronic level structure, we introduce a model based on a simple energy level scheme that includes two transitions (with temperature independent lifetimes $1/\Gamma_1$ and $1/\Gamma_2$) between common s-orbital electron levels and closely-spaced s- and p-orbital hole states. The s- and p-orbital hole levels have the same total angular momentum $j = 3/2$ and the same 8-fold degeneracy (including spin) [29]. The radiative decay rate, $\Gamma = \tau_{rad}^{-1}$, at temperature T can be calculated from the thermal population of the two transitions using Boltzmann statistics [15]:

$$\Gamma = \frac{\Gamma_1 + \Gamma_2 e^{-\Delta E/k_B T}}{1 + e^{-\Delta E/k_B T}}, \qquad \text{Eq. 1}$$

in which ΔE is the energy difference between the two hole levels. Considering this two-transition model, the critical temperature between the LT and RT regimes indicates the energy difference ΔE. We obtain the size dependence of the transition rates $\Gamma_1$ and $\Gamma_2$ and the energy splitting between the lowest energy hole levels by fitting the temperature-dependent lifetimes of Fig. 2(b) with eq. 1 [Fig. 3]. For the two larger sizes the model determines a finite transition rate, $\Gamma_1$, for the ground state transition and a nearly zero transition rate, $\Gamma_2$, for the excited state transition. Hence, in this size range the lowest energy transition is allowed and includes s-symmetry hole levels. The excited state transition is forbidden and includes p-orbital hole levels. In contrast, we find that $\Gamma_1 < \Gamma_2$ for the smallest size NC sample, which indicates that the forbidden ground state transition involves p-orbital hole levels.

Surprisingly, the energy splitting ΔE varies slowly in the measured size range, where significant changes in $\Gamma_1$ and $\Gamma_2$ occur [Fig. 3(b)]. Such a behavior is in contrast to a level crossing of two states with different emission probabilities that would result in a much stronger size dependence of ΔE. However, it can be explained by an anti-crossing of the s- and p-orbital hole levels with a characteristic minimal energy separation of ~6 meV. Such a level anti-crossing can be explained by deviations of the NC shape from the ideal spherical shape. In the anti-crossing region the s- and p-hole levels mix and the resulting hole levels are not of pure s- or p-orbital symmetry anymore. Therefore, none of the transitions that involve the lowest hole states are orbital forbidden in NCs around 2 nm radii and non-zero transition rates are associated with both lowest energy transitions [Fig. 3(b)]. While this is the first experimental account of anti-crossing of the lowest energy hole states in NCs with implications on their emission properties, anti-crossings of higher energy hole states in CdSe NCs have been measured previously [33].

The discussed $\tau_{rad}$ and associated electronic level structure can directly affect the emission efficiency of CdS NCs. Long $\tau_{rad}$ give excited electron-hole pairs more time to decay through non-radiative decay channels, thereby reducing the emission efficiency. One such decay channel is the relaxation to surface defect states that give rise to spectrally broad, low-energy photoluminescence [Fig. 3(c)]. Indeed, we measured stronger defect state emission from *small* CdS NCs that have longer $\tau_{rad}$ related to their lowest energy forbidden transition. In contrast, since the emission of *large* CdS NCs is associated with an allowed transition, it is less affected by detrimental non-radiative decay channels and the defect state emission is weaker [Fig. 3(c)].

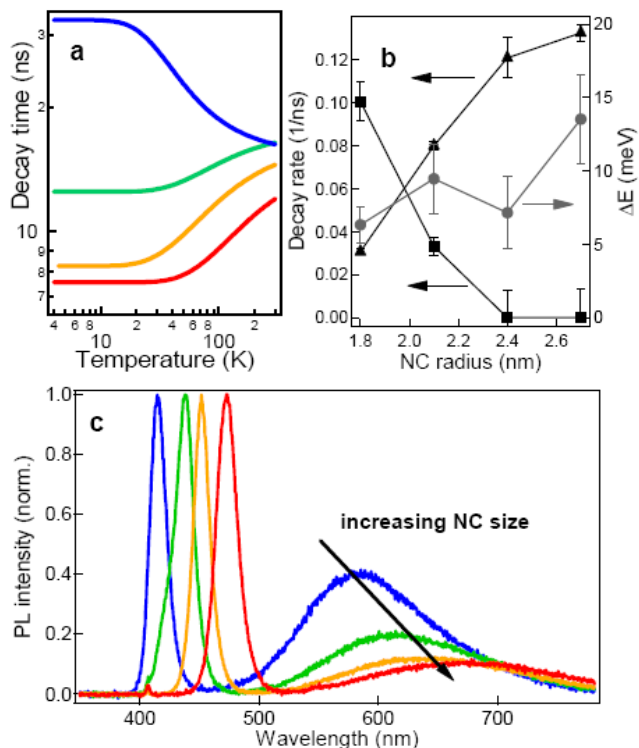

**Fig. 3:** (a) Temperature dependent lifetimes based on the two-transition model with parameters adjusted to fit the experimental data in Fig. 2(b) (blue: 1.8 nm NCs; green: 2.1 nm NCs; orange: 2.4 nm NCs; red: 2.7 nm NCs). (b) Parameters obtained in (a): transition rates of the ground, $\Gamma_1$ (triangles), and first excited, $\Gamma_2$ (square), transition and the energy spacing $\Delta E$ (circles). (c) Band edge and defect state emission (blue: 1.8 nm NCs; green: 2.1 nm NCs; orange: 2.4 nm NCs; red: 2.7 nm NCs).

In conclusion, we successfully discussed time-resolved PL experiments of CdS NCs with different sizes in the 4 - 300 K temperature range by considering two transitions between s-orbital electron and s- and p-orbital hole levels. Our data indicate that the hierarchy of the s- and p-orbital hole levels in CdS NCs depends on the NC size and switches around the NC radius of 2 nm. Uncommon for many NC systems, even at low temperatures of 4 K (0.34 meV) the radiative decay in *large* CdS nanocrystals is governed by an allowed transition that invokes an s-orbital hole state. Moreover, we concluded that the energy separation between the s- and p-orbital hole levels is always finite, indicating an anti-crossing behavior that is associated with a hole level mixing and manifested as a redistribution of oscillator strengths between the two lowest energy transitions. We expect that excitonic fine structure effects that were not taken into account here might become more prominent at temperatures below 4K and in the presence of magnetic fields. Our results contribute to the fundamental understanding of the electronic level structure and the emission properties of NCs, specifically such made of semiconductors with small spin-orbit coupling. Radiative lifetimes, as reported in this work for CdS NCs, affect the emission efficiency of NCs and, therefore, are relevant for the implementation of NCs into optoelectronic applications.

This work was supported by the NSF grant DMR-0531171 and B.Y. acknowledges an NSF-EPSCoR fellowship. The authors appreciated stimulating discussions with Al. L. Efros.

* achermann@physics.umass.edu